\documentclass[fleqn,usenatbib]{mnras}

\usepackage[T1]{fontenc}
\usepackage{ae,aecompl}

\usepackage{graphicx}	% Including figure files
\usepackage{amsmath}	% Advanced maths commands
\usepackage{amssymb}	% Extra maths symbols

\usepackage{xspace}
\usepackage[nolist]{acronym}
\usepackage{graphicx}
\usepackage{subfigure}
\providecommand{\acrolowercase}[1]{\lowercase{#1}}

\begin{acronym}
\acro{TPR}[TPR]{True Positive Rate}
\acro{FPR}[FPR]{False Positive Rate}
\acro{BNS}[BNS]{binary neutron star}
\acro{NSBH}[NSBH]{neutron star-black hole}
\acro{MDC}[MDC]{Mock Data Challenge}
\acro{EoS}[EoS]{equation of state}
\acro{NS}[NS]{neutron star}
\acro{GW}[GW]{gravitational wave}
\acro{LIGO}[LIGO]{Laser Interferometer \acs{GW} Observatory}
\acro{aLIGO}[aLIGO]{Advanced \acs{LIGO}}
\acro{AdVirgo}[AdVirgo]{Advanced Virgo}
\acro{O4}[O4]{\ac{aLIGO}'s \citep{aLIGO}, \ac{AdVirgo}'s \citep{adVirgo} and KAGRA's \citep{KAGRA:2020tym} fourth observing run}
\acro{LVK}[LVK]{LIGO-Virgo-KAGRA}
\acro{O1}[O1]{\ac{aLIGO}'s first observing run}
\acro{O2}[O2]{\ac{aLIGO}'s and \ac{AdVirgo}'s second observing run}
\acro{O3}[O3]{\ac{aLIGO}'s and \ac{AdVirgo}'s third observing run}
\acro{oLIB}[\acrolowercase{o}LIB]{Omicron+\acl{LIB}}
\acro{KAGRA}[KAGRA]{KAmioka GRAvitational\nobreakdashes-wave observatory}
\acro{GraceDB}[\texttt{GraceDB}]{GRAvitational-wave Candidate Event DataBase}
\acro{LLAI}[LLAI]{Low-Latency Alert Infrastructure}
\acro{CBC}[CBC]{compact binary coalescence}
\acro{BBH}[BBH]{binary black hole}
\acro{FAR}[FAR]{false alarm rate}
\acro{GstLAL}[GstLAL]{GStreamer LIGO Scientific Collaboration Algorithm Library}
\acro{SVD}[SVD]{singular value decomposition}
\acro{SNR}[SNR]{signal\nobreakdashes-to\nobreakdashes-noise ratio}
\acro{FGMC}[FGMC]{Farr, Gair, Mandel and Cutler formalism of Poisson Mixture Model}
\acro{MBTA}[MBTA]{Multi-Band Template Analysis}
\acro{SPIIR}[SPIIR]{Summed Parallel Infinite Impulse Response}
\acro{GPU}[GPU]{Graphical Processing Units}
\acro{cWB}[\acrolowercase{c}WB]{Coherent WaveBurst}
\acro{GCN}[GCN]{General Coordinates Network}
\acro{SCiMMA}[SCiMMA ]{Scalable CyberInfrastructure for Multi-Messenger Astrophysics}
\acro{BAYESTAR}[\texttt{BAYESTAR}]{BAYESian TriAngulation and Rapid localization}
\acro{FITS}[FITS]{Flexible Image Transport System}
\acro{HEALPix}[HEALP\acrolowercase{ix}]{Hierarchical Equal Area isoLatitude Pixelization}
\acro{MOC}[MOC]{Multi-Order Coverage}
\acro{EM}[EM]{electromagnetic}
\acro{NS}[NS]{neutron star}
\acro{BH}[BH]{black hole}
\acro{LLOID}[LLOID]{Low Latency Online Inspiral Detection}
\acro{IIR}[IIR]{infinite impulse response}
\acro{ToO}[ToO]{target of opportunity}
\acro{PE}[PE]{Parameter Estimation}
\acro{Bilby}[Bilby]{Bayesian Inference Library for \acs{CBC} \acs{GW} signal}
\acro{RAVEN}[RAVEN]{Rapid, on-source VOEvent Coincident Monitor}
\acro{NMMA}[\texttt{NMMA}]{The Nuclear Multimessenger Astronomy Framework}

\end{acronym}

\usepackage{newtxtext,newtxmath}
\title[]{What to expect: kilonova light curve predictions via equation of state marginalization}

\author[Toivonen et al.]{Andrew Toivonen$^{1}$\thanks{E-mail: toivo032@umn.edu}, Gargi Mansingh$^{1,2,3}$, Holton Griffin$^{1}$, Armita Kazemi$^{1}$, Frank Kerkow$^{1}$,
\newauthor{Stephen K. Mahanty$^{1}$, Jacob Markus$^{1}$, Seiya Tsukamoto$^{1}$, Sushant Sharma Chaudhary$^{4}$,}
\newauthor{Sarah Antier$^{5}$, Michael W. Coughlin$^{1}$, Deep Chatterjee$^{6}$, Reed Essick$^{7}$, Shaon Ghosh$^{8}$,}
\newauthor{Tim Dietrich$^{9,10}$, Philippe Landry$^{7}$}
\\
$^{1}$School of Physics and Astronomy, University of Minnesota, Minneapolis, Minnesota 55455, USA\\
$^{2}$Department of Physics, American University, 3501 Nebraska Ave NW, Washington, DC 20016\\
$^{3}$Nicholas \& Lee Begovich Center for Gravitational-Wave Physics \& Astronomy, California State University, Fullerton, 800 N State College Blvd, Fullerton, CA 92831\\
$^{4}$Institute of Multi-messenger Astrophysics and Cosmology,
Missouri University of Science and Technology, Physics Building, 1315 N. Pine St., Rolla, MO 65409, USA \\
$^{5}$Observatoire de la C\^ote d'Azur, Universit\'e C\^ote d'Azur, Boulevard de l'Observatoire, 06304 Nice, France \\
$^{6}$MIT Kavli Institute for Astrophysics and Space Research 77 Massachusetts Avenue, McNair Building (MIT Building 37)
Cambridge, MA 02139\\
$^{7}$Canadian Institute for Theoretical Astrophysics, 60 St. George St, Toronto, Ontario M5S 3H8, Canada\\
$^{8}$Montclair State University, 1 Normal Ave, Montclair, NJ 07043\\
$^{9}$Institut f\"ur Physik und Astronomie, Universit\"at Potsdam, Haus 28, Karl-Liebknecht-Strasse 24/25, 14476, Potsdam, Germany \\
$^{10}$Max Planck Institute for Gravitational Physics (Albert Einstein Institute), Am M\"uhlenberg 1, Potsdam 14476, Germany
}

\date{Accepted XXX. Received YYY; in original form ZZZ}

\pubyear{\the\year{}}

\begin{document}
\label{firstpage}
\pagerange{\pageref{firstpage}--\pageref{lastpage}}
\maketitle

\begin{abstract}
Efficient multi-messenger observations of gravitational-wave candidates from \ac{CBC} candidate events rely on data products reported in low-latency by the International Gravitational-wave Network (IGWN). 
While data products such as \texttt{HasNS}, the probability of at least one neutron star, and \texttt{HasRemnant}, the probability of remnant matter forming after merger, exist, these are not direct observables for a potential kilonova. 
Here, we present new kilonova light curve and ejecta mass data products derived from merger quantities measured in low latency, by marginalizing over our uncertainty in our understanding of the neutron star \ac{EoS} and using measurements of the source properties of the merger, including masses and spins.
Two additional types of data products are proposed.
The first is the probability of a candidate event having mass ejecta ($m_{\mathrm{ej}}$) greater than $10^{-3} M_\odot$, which we denote as \texttt{HasEjecta}.
The second are $m_{\mathrm{ej}}$ estimates and accompanying \texttt{ugrizy} and \texttt{HJK} kilonova light curves predictions produced from a surrogate model trained on a grid of kilonova light curves from \texttt{POSSIS}, a time-dependent, three-dimensional Monte Carlo radiative transfer code.
We are developing these data products in the context of the IGWN low-latency alert infrastructure, and will be advocating for their use and release for future detections.

\end{abstract}


%%%%%%%%%%%%%%%%%%%%%%%%%%%%%%%%%%%%%%%%%%%%%%%%%%

%%%%%%%%%%%%%%%%% BODY OF PAPER %%%%%%%%%%%%%%%%%%

\section{Introduction}
\label{sec:intro}

The combined detection of the kilonova AT2017gfo \citep{CoFo2017,SmCh2017,AbEA2017f} and gravitational-wave (GW) observations resulting from the binary neutron star merger GW170817 \citep{AbEA2017b} has led to immense interest in the field of multi-messenger astrophysics. Kilonovae are short lived astrophysical transients that may result from either \ac{BNS}, or a \ac{NSBH} mergers and are of particular astrophysical interest due to the fact that kilonovae are expected to be sites of r-process nucleosynthesis, through which it is theorized heavy elements can be produced. The radioactive decay and interactions of these r-process elements are what powers the kilonova emission we hope to observe~\citep{LaSc1974,LiPa1998,MeMa2010,KaMe2017}. While GW170817 led to breakthroughs in nuclear astrophysics \citep{Margutti:2017cjl, Smartt:2017fuw, KaNa2017, Kasen:2017sxr, Watson:2019xjv}, cosmology \citep{Abbott:2017xzu, Coughlin:2019vtv, Dietrich_2020}, and tests of General Relativity \citep{Ezquiaga:2017ekz, Baker:2017hug, Creminelli:2017sry}, much remains to be learned about these rare events, including the diversity of their intrinsic parameters, their emission, and the heavy elements produced by r-process nucleosynthesis in these mergers.

\ac{O4} is underway as of May 23, 2023\footnote{\url{https://observing.docs.ligo.org/plan}}, and the search for gravitational wave events and their counterparts \citep{Abbott_2020} has resumed. Searches for kilonovae are challenging due to the fact that they are short lived events, can be relatively faint, and may not be well localized. Sky localizations for gravitational wave events can span $\approx100-10,000 \ \mathrm{deg^2}$ \citep{Rover2007a, Fair2009,Fair2011,Grover:2013,WeCh2010,SiAy2014,SiPr2014,BeMa2015,EsVi2015,CoLi2015,KlVe2016}. However, in spite of these challenges, it is imperative to locate the transient as quickly as possible in the hope of observing the peak of emissions. A number of wide-field survey telescopes are used to try and cover these large sky localizations, such as: the Panoramic Survey Telescope and Rapid Response System (Pan-STARRS) \citep{MoKa2012}, Asteroid Terrestrial-impact Last Alert System (ATLAS) \citep{ToDe2018}, the Zwicky Transient Facility (ZTF) \citep{Bellm2018,Graham2018,Ahumada:2024qpr}, and in the near future BlackGEM \citep{BlGr2015}, the Large Synoptic Survey Telescope (LSST) \citep{LSST:2008ijt}, the Nancy Grace Roman Space Telescope\footnote{\url{https://roman.gsfc.nasa.gov/}} \citep{Andreoni:2023xlv}, and Ultraviolet Explorer (UVEX) \citep{Kulkarni:2021tit}.
While GW178017 is the only \ac{CBC} merger event that has provided us with joint observations of a GW signal, kilonova, and short gamma-ray burst (sGRB), it is also potentially possible to identify kilonovae associated with sGRBs \citep{Tanvir:2013pia, AsCo2018, Jin:2019uqr, Rastinejad:2022zbg}, or even serendipitously in survey operations \citep{Andreoni:2021ykx, Almualla:2020ybm}.

These searches are aided by source classification efforts, which we can use to determine the origin of \ac{GW} events \citep{Chatterjee:2019avs, Berbel:2023vug}. In addition, there have been multiple efforts to simulate \ac{GW} detections and constrain rates for \ac{O4} and beyond \cite{Petrov:2021bqm, Colombo:2022zzp, Kiendrebeogo:2023hzf}. With \ac{O4} underway, we hope for additional \ac{BNS} detections and follow-up opportunities. These observing scenarios simulations tell us what we can expect to observe, and can even help us constrain poorly measured parameters, like the inclination angle, as discussed in Sec. \ref{subsec:inclination}.
The parameters of the binary before merger, such as the mass ratio and the masses of the objects involved, along with the \ac{EoS}, can help predict the mass ejected from the merger and the light curves associated with a possible kilonova 
\citep{BaBa2013,PiNa2013,AbEA2017b,BaJu2017,DiUj2017,RaPe2018}.
The relationship between light curves and binary parameters can also be used to place constraints on the character of the progenitor systems and the mass ejected \citep{CoDi2017,SmCh2017,Coughlin_2018miv, Bulla:2022mwo}. 

Predicting whether we can expect to see a kilonova, how bright that kilonova may be, and where it will be localized in the sky are all crucial for astronomers who work on follow-up of \ac{GW} events. Currently, the sky localization is provided through sky maps, produced by \texttt{BAYESTAR} \citep{SiPr2016} and \texttt{Bilby} \citep{Ashton:2018jfp}, These efforts are intended to allow for more informed and efficient follow-up searches. 
There are also existing properties, \texttt{HasNs} and \texttt{HasRemnant} \citep{Chatterjee:2019avs}, which are provided by the \texttt{EM-bright} pipeline and encode the probability that there will be at least one \ac{NS} and non-zero $m_{\mathrm{ej}}$, respectively, by providing estimates of the amount of $m_{\mathrm{ej}}$ produced and the $M_{AB}$ of the kilonova.

In this paper, we propose two new data products, focused on observable properties directly tied to kilonovae, partially building on the work of \citep{Stachie_2021}. We use similar ejecta fits and also define a quantity \texttt{HasEjecta} to describe the likelihood of having a significant amount of $m_{\mathrm{ej}}$. One noticeable difference is our focus on using parameter estimation, instead of template based estimates. The first data product is the probability of a candidate event having $m_{\mathrm{ej}}$ greater than $10^{-3} M_\odot$; we denote this quantity as \texttt{HasEjecta}. 
This quantity is useful for establishing the likelihood of a kilonova counterpart, regardless of brightness. The $10^{-3} M_\odot$ cutoff was chosen for three reasons, (i) our kilonova surrogate models were not trained on grid points below this value, (ii) ejecta fits can have large uncertainties that dominate at low values, and (iii) this provides a reasonable floor estimate for the total $m_{\mathrm{ej}}$ capable of producing an observable kilonova.
To directly support electromagnetic counterpart searches, we also propose to provide predicted \texttt{ugrizy} and \texttt{HJK} light curves.
To do so, we use a surrogate model based on the time-dependent, three-dimensional Monte Carlo radiative transfer code \texttt{POSSIS} \citep{Bul2019}. Using \texttt{POSSIS}, we produce kilonova light curves by simulating packets of photons released by the radioactive decay of r-process nuclei in a kilonova. \texttt{POSSIS} takes input in the form of a model that defines densities, compositions, and geometry of the ejecta, and outputs light curves and spectra as a function of inclination angle. \texttt{POSSIS} simulations are computationally expensive, so a grid of kilonova light curves are computed and then used to train a surrogate light curve model with \ac{NMMA}\footnote{\url{https://github.com/nuclear-multimessenger-astronomy/nmma}}  \citep{Pang:2022rzc}. \ac{NMMA} is a nuclear physics and cosmology library used for analysis of \ac{BNS} and \ac{NSBH} systems, as well as potential counterparts. To create the light curve model, \ac{NMMA} starts with a grid of \texttt{POSSIS} simulations across input parameters, and uses a neural network to interpolate between that grid and generalize to arbitrary parameters. For this work, we use the light curve model referred to as \texttt{Bu2019lm} \citep{Bul2019}, which depends on the dynamical mass ejecta ($m^{\mathrm{dyn}}_{\mathrm{ej}}$), wind ejecta $m^{\mathrm{wind}}_{\mathrm{ej}}$, inclination angle ($\theta$), and the opening angle $\phi$, as will be discussed more later on.

Sec.~\ref{sec:modeling} provides the workflow for producing the proposed data products from what is available in low latency from IGWN searches.
We present performance for the proposed data products in Sec.~\ref{sec:analysis}, while
Sec.~\ref{sec:conclusion} provides our conclusions and plans for future research.

%\subsection{Kilonova searches}
%\subsection{Constraints on kilonovae}
%\subsection{Impact of kilonovae observations}

\section{Modeling}
\label{sec:modeling}

\begin{figure*}
 \includegraphics[width=\textwidth]{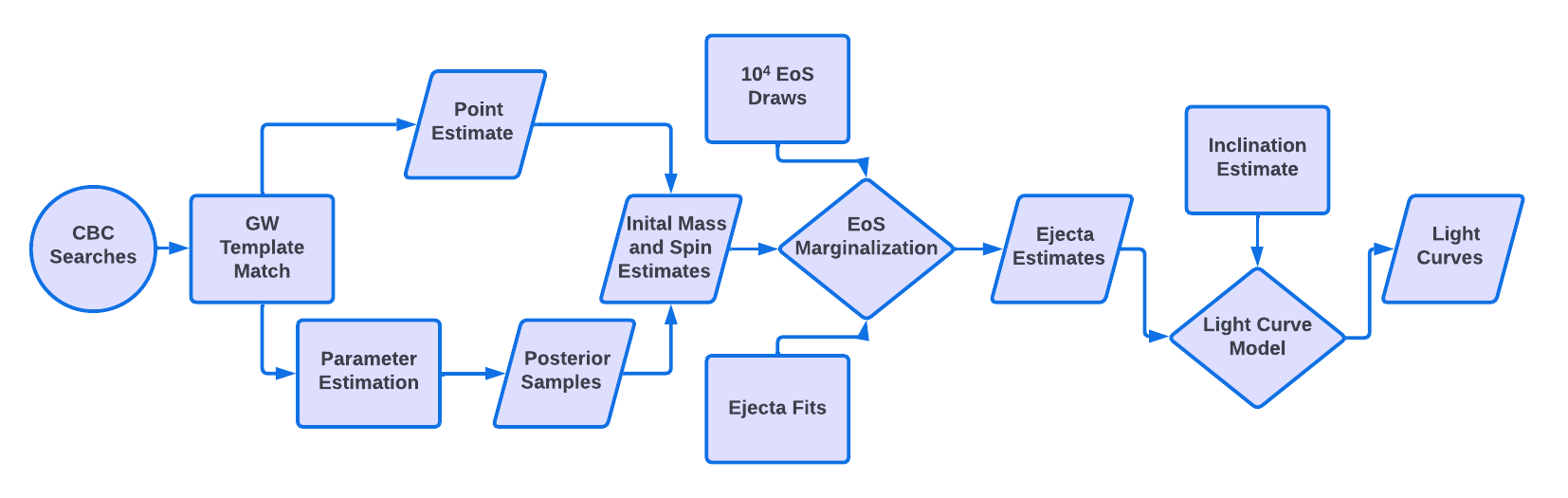}
  \caption{Workflow for the light curve and ejecta mass data products. We start with initial mass and spin estimates from a \ac{GW} candidate as input, then marginalize over a number of \ac{EoS} realizations per sample and apply mass ejecta fits. Finally, we use our light curve model to calculate light curves after drawing inclination angles if needed.}
 \label{fig:workflow}
\end{figure*}

\subsection{Workflow overview}
Fig.\ref{fig:workflow} provides an overview of the workflow of our kilonova ejecta and light curve predictions. We start with a \ac{GW} detection by one of the \ac{CBC} searches, and then use initial mass and spin estimates from either the point estimate, or parameter estimation, as discussed in Sec. \ref{subsec:chirp}. The point estimate has high uncertainties, but is also available in low-latency at the time the alert is sent out. Parameter estimation is more accurate, and covers a range of mass and spin values, but is not available until $\sim$ hours after merger. From there, we marginalize over \ac{EoS} realizations and use the ejecta fits covered in Sec. \ref{subsec:fits} to calculate the $m_{\mathrm{ej}}$. Each sample is run with a set number of random \ac{EoS} realizations to effectively cover the parameter space. Finally, adding in an inclination distribution and opening angle, covered in Sec. \ref{subsec:inclination} and using our kilonova model covered in Sec. \ref{subsec:lightcurves}, we generate kilonova light curves. 

%\subsection{Compact binary merger distributions} 

%Due to the fact that we only have one joint kilonova and \ac{GW} detection, we do not have a known sample set to inform parameters of such events. At times, we must make use of theoretical population level distributions of compact objects. We can make ejecta and light curve predictions of compact binary merger events from a variety of sources, including theoretical component mass distributions, distributions based on the GWTC-3 catalog \citep{KAGRA:2021vkt}, and from parameter estimation of Mock Data Challenge (MDC) injections \citep{Meacher:2015iua}. The events based on GWTC-3 we used were simulated by Coughlin and Farah \citep{michael_w_coughlin_2022_7026209}. In addition, we can use theoretical distributions for merger populations of component masses in order to make predictions. For neutron stars we use the Farrow \citep{Farrow_2019} and Alsing \citep{Alsing_2018} distributions, and for black holes we use the Zhu \citep{Zhu_2021} distribution. We use a variety of these distributions to generate the plots in this paper and analyze our results. In the following sections we examine the merger parameters that affect our predictions.

\subsection{Chirp mass and mass ratio}
\label{subsec:chirp}

The chirp mass and mass ratio of a compact object merger are important quantities for characterization of a merger event. The chirp mass is the simplest merger quantity to measure from a \ac{GW} signal, due to its relationship with the frequency evolution of the signal. Chirp mass, $m_{\rm chirp}$, is defined as:
\begin{equation}
    m_{\rm chirp} = \dfrac{(m_1 m_2)^{3/5}}{(m_1+m_2)^{1/5}}
\end{equation}
where $m_1$ is the primary, or larger, mass, while $m_2$, the secondary, or smaller, mass. The mass ratio, $q$, is much more difficult to measure accurately from \ac{GW} data, and is the ratio between the component masses. 
\begin{equation}
    q = \dfrac{m_2}{m_1}
\end{equation}

Estimates of the $q$ can be made using the phase of the \ac{GW} signal, but they include significant error. The chirp mass is recovered quite accurately by parameter estimation in most cases, however, there may still be a wide range of possible component mass values due to the poorly constrained mass ratios, meaning the source classification cannot always be clearly ascertained. 

We have two possible options for our initial mass and spin estimates: (i) the  mass estimates produced by \texttt{Bilby} \citep{Ashton:2018jfp}, an automated Bayesian parameter estimation analysis library, and (ii) the point estimate from the \ac{CBC} template match, e.g. \texttt{GstLAL} \citep{Messick:2016aqy, Tsukada:2023edh, Ewing:2023qqe}, \texttt{MBTA} \citep{Andres:2021vew}, \texttt{PyCBC} \citep{Nitz:2018rgo, DalCanton:2020vpm}, \texttt{SPIIR} \citep{Luan:2011qx, Hooper:2011rb}. While parameter estimation is the starting point with the most precise measurements, given it provides (i) the most accurate estimate of the true source parameters and (ii) uncertainties, it takes $\sim$ hours to be completed post-merger. If a low-latency prediction is desired, as may benefit searches for counterparts in the hours after merger, we must use the point estimate as a starting point. There are also promising prospects of real-time machine learning based parameter estimation, which may be available in the future \citep{Dax:2021tsq}, but for now we must rely on the point estimate. 
In this case, we can use the point estimate, or take the chirp mass from the point estimate and draw a mass ratio consistent with that chirp mass value from population level distributions \citep{michael_w_coughlin_2022_7026209, Kiendrebeogo:2023hzf} informed by GWTC-3 catalogue fits \citep{KAGRA:2021vkt}. From this chirp mass and mass ratio, we compute the component masses that will be inputs for the simulated ejecta quantities and light curves (see below). For the analysis of \ac{MDC} \citep{Chaudhary:2023vec} events to follow, we use parameter estimation as the starting point for the reasons outlined above.

%We calculate ejecta and light curve estimates in the source frame, so if needed the detector frame masses are shifted to the source frame using the redshift. 

%Figure \ref{fig:mej_inj} demonstrates an example of an Mock Data Challenge (MDC) \citep{Meacher:2015iua} injection compared to the marginalized predictions. We compare the injected light curve, computed using the \ac{EoS} SLy \citep{Chabanat:1997qh, Chabanat:1997un} (or: https://arxiv.org/abs/astro-ph/0111092), to the \ac{EoS} marginalized prediction using the chirp mass from the posterior samples, and drawing the mass ratio as explained above. This reliably produces a histogram of ejecta samples consistent with the ejecta of the injection. 

%A wider mass ratio prior is being re-implemented for parameter estimation of MDC events, so we plan to test these samples again and compare the results using the PE mass prior versus our values drawn from GWTC-3 fits. In the future we hope to be able to use the mass ratio values provided for O4 events. 

\subsection{Equation of state and spin distributions}

\begin{figure}
 \includegraphics[width=3.5in]{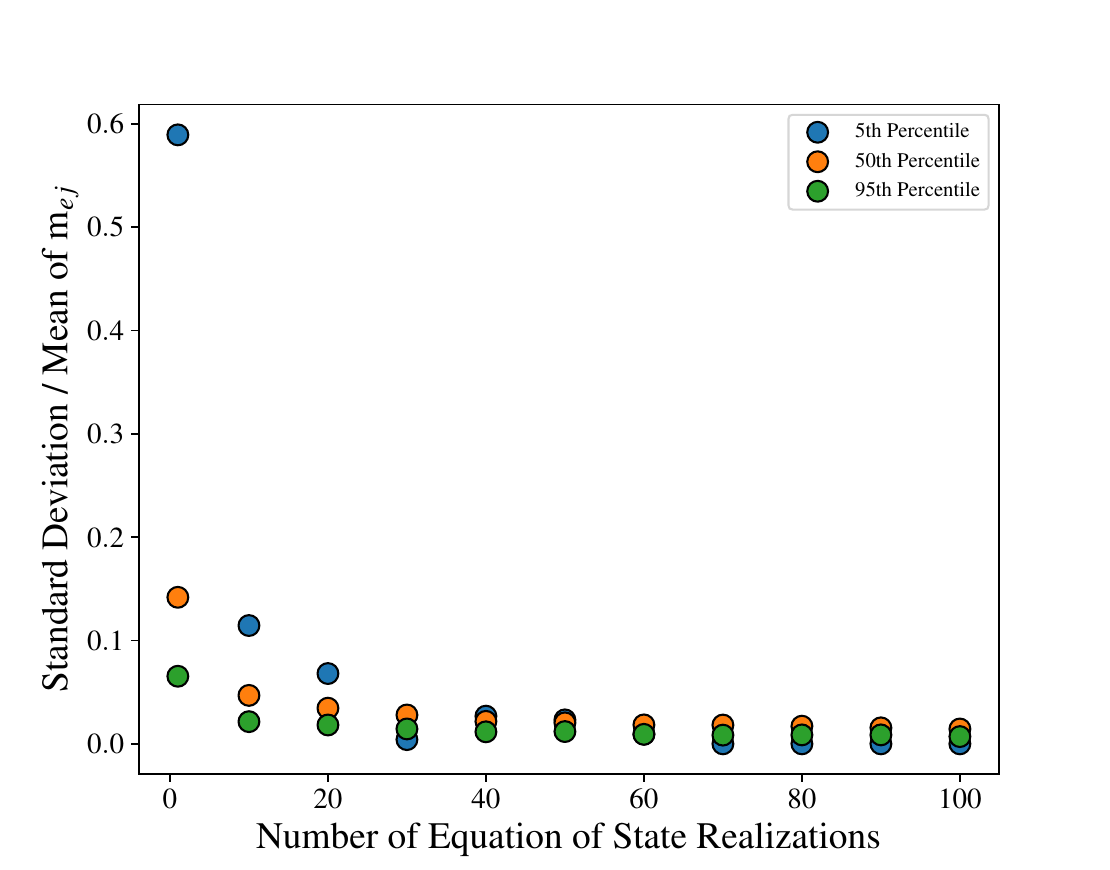}
  \caption{Comparison of relative error between 100 realizations of predictions based on parameter estimation for a single \ac{MDC} event as a function of the number of \ac{EoS} realizations per sample. We find that the error falls off asymptotically as the number of \ac{EoS} realizations increasing, and that there are diminishing returns past ~50 \ac{EoS} realizations, where the error is less than a few percent.} 
 \label{fig:EoS}
\end{figure}

The \ac{NS} \ac {EoS} impacts the amount of mass expected to be ejected from a \ac{CBC} merger as it influences how the \ac{NS} is tidally disrupted. 
While the \ac{NS} \ac{EoS} remains unknown, there are certain popular \ac{EoS} in the literature useful for comparisons. In the following, when we need a single \ac{EoS} to compare to, we will use the SLy \citep{ChBo1998} \ac{EoS} for comparisons to our marginalized range of predictions. This is to provide a singular point of reference for comparison, and to ensure that our predictions are reasonable. The SLy \ac{EoS} was chosen as it has support from mass-radius posteriors from GW170817
\citep{LIGOScientific:2018cki}.

Within our pipeline, due to the uncertainty in our understanding of the \ac{NS} \ac{EoS}, we marginalize over a number of \ac{EoS} realizations per sample in order to produce our ejecta estimates. For each component mass pair, we draw a specified number of \ac{EoS} realizations from a set of $10^{4}$ equally weighted \ac{EoS} realizations in order to cover the parameter space. The \ac{EoS} realizations are drawn from a Gaussian-process posterior \citep{Legred:2021hdx, legred_isaac_2022_6502467} conditioned on a radio pulsar mass measurement \citep{Zhao:2015tra} and gravitational-wave mass and tidal deformability measurements \citep{AbEA2017b, LIGOScientific:2020aai}. For the purpose of our workflow, each sample is run with a different subset but same number of \ac{EoS} realizations. This number of realizations is a variable within the workflow and can be set as desired. To establish the number sufficient for consistent results, we ran a series of tests; Fig.~\ref{fig:EoS} shows how the relative error, defined here as the standard deviation over the median $m_{\mathrm{ej}}$, varies for different numbers of \ac{EoS} realizations. The standard deviation and mean was found across 100 realizations of the predictions for each number of \ac{EoS} realizations. We find that 50 \ac{EoS} realizations reduces the error to less than a few percent, giving us more than sufficient consistency between runs while still being computationally feasible. Beyond this point improvements are negligible. 50 \ac{EoS} realizations also ensures that even in the case where only a small fraction of the posterior samples produce significant amounts of $m_{\mathrm{ej}}$, we will still adequately cover the parameter space. If running on a point estimate or similar small sample size, all $10^{4}$ samples can be employed.
 
The \ac{NS} and \ac{BH} spins may also impact the amount of $m_{\mathrm{ej}}$. As will be covered in Sec.~\ref{subsec:fits}, the \ac{BNS} ejecta fits employed do not depend on spin, while the \ac{NSBH} depend on the total effective spin $\chi_{\rm eff}$:
\begin{equation}
    \chi_{\mathrm{eff}} = \dfrac{\chi_{\rm 1,z}/m_1 + \chi_{\rm 2,z}/m_2}{m_1 + m_2}
\end{equation}

This means that our data products will be unchanged by varying spin values for \ac{BNS} events but impacted for \ac{NSBH} events. For simulated events from the \ac{MDC} that we will compare to in Sec.~\ref{sec:analysis}, the spin distributions are injected uniform in magnitude and isotropic in orientation, with magnitudes up to $0.4$ and $1$ for \ac{NS} and \ac{BH} respectively \citep{Chaudhary:2023vec}. Just as for candidate \ac{GW} events, parameter estimation provides spin estimates alongside the component masses, which can then be used for ejecta predictions.

%The spins used in the analysis
%Our BNS ejecta fits do not depend on spin, so the injected spin has no influence over our BNS predictions. In the \ac{NSBH} case, the ejecta fits depend on the total effective spin, and so we have the ability to marginalize over a range of spin values. \ac{NSBH} mergers with high spin are more likely to produce significant ejecta, however, and we will be marginalizing over the effective spin for \ac{NSBH} mergers in the future.  

\subsection{Dynamical and disk wind ejecta from fits: BNS and NSBH}
\label{subsec:fits}

\begin{figure*}
 \includegraphics[width=\textwidth]{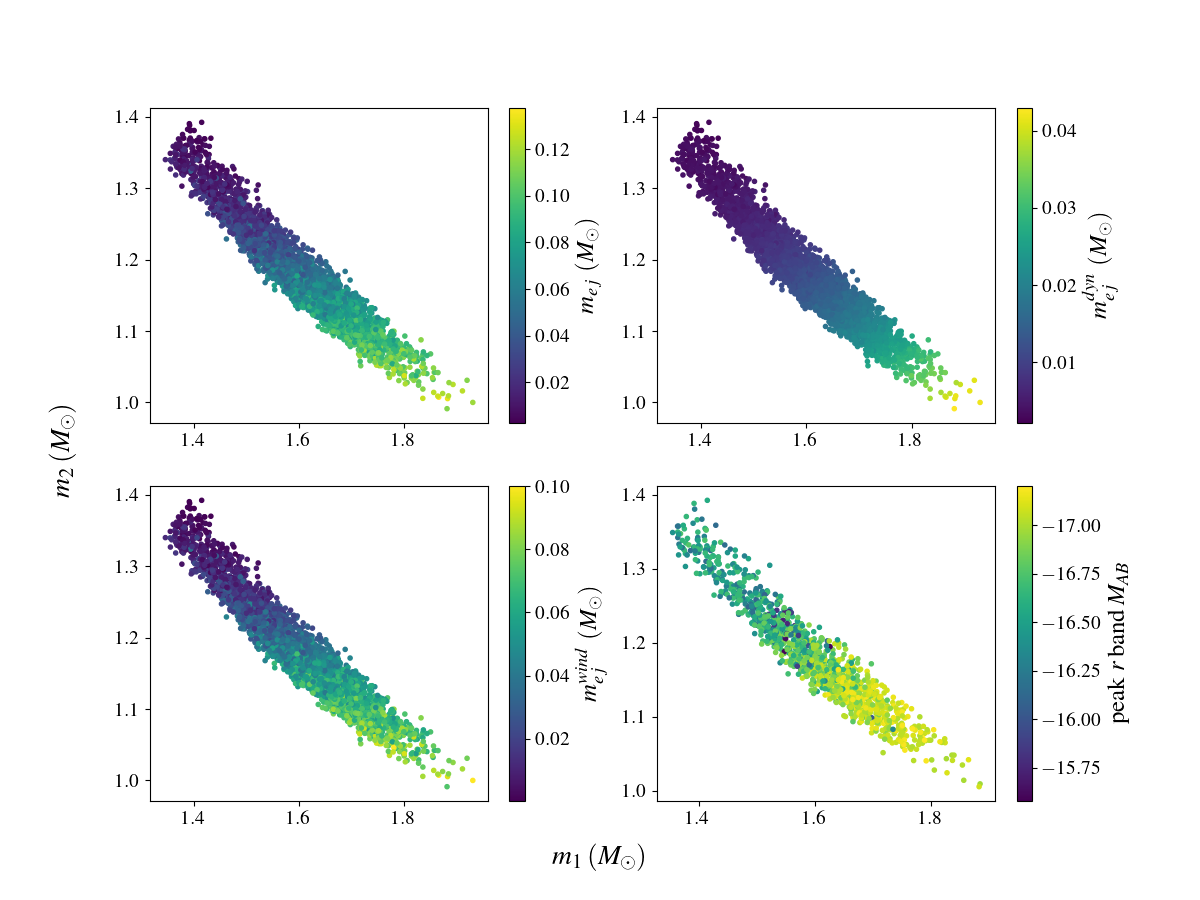}
  \caption{Scatter plots of $m_{\mathrm{ej}}$ and light curve predictions using \ac{EoS} marginalization and parameter estimation from a $m_1 = 1.40 M_\odot$, $m_2 = 1.34 M_\odot$ injection. On the upper left we have $m_{\mathrm{ej}}$, upper right $m^{\mathrm{dyn}}_{\mathrm{ej}}$, lower left $m^{\mathrm{wind}}_{\mathrm{ej}}$, and on the lower right the peak $r$-band $M_{AB}$. We see a strong correlation between all four quantities with some slight variations. }
 \label{fig:dyn_wind}
\end{figure*}

Having covered the initial merger parameters, we shift our focus to the resulting merger products. The total $m_{\mathrm{ej}}$ is calculated in two components: the dynamical ejecta, $m^{\mathrm{dyn}}_{\mathrm{ej}}$, and the disk wind ejecta, $m^{\mathrm{wind}}_{\mathrm{ej}}$.
The first component, $m^{\mathrm{dyn}}_{\mathrm{ej}}$ is produced as the extreme tidal forces of the inspiral tidally deforms and rips mass from a neutron star \citep{Metzger:2016pju}. 
The other, $m^{\mathrm{wind}}_{\mathrm{ej}}$, is produced by matter that is ejected from the accretion disk of the merger by energetic outflows of particles, sometimes referred to as particle ``winds'' \citep{Metzger:2016pju}. 
Estimates of $m_{\mathrm{ej}}$ are found from fit by combining the source properties with the \ac{EoS} used. The \ac{EoS} provides tables of mass, radius, and tidal deformability information \citep{legred_isaac_2022_6502467} which can be used to determine compactness needed for the fits. 
When making predictions, it is possible to (i) marginalize over \ac{EoS} in order to effectively cover the uncertainty in this parameter space, or (ii) pick a single named \ac{EoS} as input. Regardless of the \ac{EoS}, we use ejecta fits in order to to calculate both the dynamical and disk wind ejecta, which are found from separate fits for \ac{BNS} mergers and \ac{NSBH} mergers. These are found in Eqs. \ref{eq:1}-\ref{eq:8} below. We use the fits as implemented in \ac{NMMA} \citep{Pang:2022rzc}.

Moving onto the specific details of the calculations, the total $m_{\mathrm{ej}}$ is defined by \cite{Dietrich_2020} in Eq. \ref{eq:1}, where $M_{\rm disk}$ is the mass of the disk and $\zeta$ is the fraction of mass that becomes unbound from the system. Therefore $\zeta M_{\rm disk}$ is the disk wind ejecta: $m^{\mathrm{wind}}_{\mathrm{ej}}$. We use $\zeta = 0.30$ for both \ac{BNS} and \ac{NSBH} mergers (\cite{Fern_ndez_2014}; \cite{Christie_2019}; \cite{Siegel_2018}; \cite{Fern_ndez_2018}). The total $m_{\mathrm{ej}}$ is then simply the sum of the contributions from the $m^{\mathrm{dyn}}_{\mathrm{ej}}$ and $m^{\mathrm{wind}}_{\mathrm{ej}}$ components.

\begin{equation} \label{eq:1}
    m_{\rm ej} = M_{\rm ej}^{dyn} + \zeta M_{\rm disk}
\end{equation}

% confirmed correct
For \ac{BNS} mergers, we use the following fit in Eq. \ref{eq:2} from \citep{Kruger:2020gig} for $m^{\mathrm{dyn}}_{\mathrm{ej}}$:

\begin{equation} \label{eq:2}
    \dfrac{M^{dyn}_{\rm ej}}{10^{-3} M_\odot} = \mathrm{max}\left(0, \left[ \dfrac{a}{C_1} + b + \left(\dfrac{m_1}{m_2}\right)^{n} + cC_1\right]m_1+[1\leftrightarrow2] \right)
\end{equation}

In these expressions, $m_1$ is the mass of the primary, or largest component mass, while $m_2$ is the secondary, or smaller mass. For \ac{BNS} both are neutron stars, while for \ac{NSBH} $m_1$ is always the black hole, while $m_2$ is the neutron star. In the same way, $C_1$ is the compactness of the primary mass, while $C_2$ is the compactness of the secondary. From the fit, the coefficients are: $a = -9.335$ $b = 114.17$, $c = -337.56$, and $n = 1.5465$. The term $[1\leftrightarrow2]$ simply refers to an addition of the same terms with indices swapped.

Then, for $m^{\mathrm{wind}}_{\mathrm{ej}}$, again for \ac{BNS} mergers, we use the following in Eq. \ref{eq:3} from \cite{Dietrich_2020}:

\begin{equation} \label{eq:3}
    \mathrm{log}_{10}\dfrac{M_{\rm disk}}{M_\odot} = \mathrm{max}\left( -3,a\left(1+b\tanh\left(\dfrac{c-m_{\rm tot}/M_{\rm threshold}}{d}\right)\right)\right)
\end{equation} 

where $a$ and $b$ are given by

\begin{equation} \label{eq:4}
    a = a_o + \delta a \cdot \xi\,, \qquad
    b = b_o + \delta b \cdot \xi\,,
\end{equation}

and $a_o$, $b_o$, $\delta a$, $\delta b$, $c$, and $d$ are all free parameters. The parameter $\xi$ is given by

\begin{equation} \label{eq:5}
    \xi = \frac{1}{2}\tanh\left(\beta \left(q-q_{\rm trans}\right)\right)\,,
\end{equation}

where $\beta$ and $q_{\rm trans}$ are free parameters and $q \equiv m_2/m_1 \leq 1$ is defined as the mass ratio. 
The best-fit model parameters are $a_o=-1.581$, $\delta a=-2.439$, $b_o=-0.538$, $\delta b=-0.406$, $c =0.953$, $d=0.0417$, $\beta=3.910$, $q_{\rm trans}=0.900$. The threshold mass $M_{\rm threshold}$ for a given \ac{EoS} is estimated by the following \citep{Agathos_2020}:
\begin{equation} \label{eq:6}
    M_{\rm threshold} = \left(2.38 - 3.606 \frac{M_{\rm TOV}}{R_{1.6}}\right) M_{\rm TOV},
\end{equation}

where $M_{\rm TOV}$ is the maximum mass of a non-spinning NS and and $R_{1.6}$ is the radius of a $1.6M_{\odot}$ NS.

%where $m_{\rm tot}$ is simply the sum of $m_1$ and $m_2$. $M_{\rm thresh}$ is the threshold mass, which defines the maximum mass of a merger remnant for which anything more massive will collapse into a black hole \citep{Agathos_2020}. For this fit, $c = 0.953$ and $d = 0.0417$, while $a$ and $b$ are not constants and depend on the mass ratio and other values, as outlined in \cite{Dietrich_2020}.

Now moving to \ac{NSBH} mergers, we use the $m^{\mathrm{dyn}}_{\mathrm{ej}}$ fit from \cite{Kruger:2020gig}, shown in Eq. \ref{eq:7}.

% confirmed correct https://github.com/nuclear-multimessenger-astronomy/nmma/blob/main/nmma/joint/conversion.py#L108

\begin{equation} \label{eq:7}
    M_{\rm dyn}(M_\odot) = m^{\rm bar}_2 \left(a_1\left(\dfrac{m_1}{m_2}\right)^{\rm n_1} \dfrac{1-2C_2}{C_2}-a_2\left(\dfrac{m_1}{m_2}\right)^{\rm n_2} \dfrac{r_{\rm ISCO}}{m_1}+a_4\right)
\end{equation}

The coefficients are: $a_1 = 0.007116$, $a_2 = 0.001436$, $a_4 = -0.02762$, $n_1 = 0.8636$, and $n_2 = 1.6840$. 

Again for \ac{NSBH}, the $m^{\mathrm{wind}}_{\mathrm{ej}}$ fit is defined by \cite{Foucart_2018} here in Eq. \ref{eq:8}.

% Correct: https://journals.aps.org/prd/pdf/10.1103/PhysRevD.98.081501 

\begin{equation} \label{eq:8}
    M_{\rm disk}(M_\odot) = m^{\rm bar}_2 \mathrm{max}\left(0,\alpha \dfrac{1-2C_2}{\eta^{1/3}}-\beta r_{\rm ISCO}\dfrac{C_2}{\eta}+\gamma\right)^\delta
\end{equation}

In Eq. \ref{eq:8}, $m^{\rm bar}_2$ refers to the baryonic mass of the secondary mass, $\eta$ is the reduced mass and is defined as $\eta=(m_1m_2)/(m_1+m_2)$, and $r_{\rm ISCO}$ is the innermost stable circular orbit of the binary. From the fit, the coefficients are: $\alpha = 0.4064$, $\beta = 0.1388$, $\gamma = 0.2551$, and $\delta = 1.7612$.

Predicted distributions of $m_{\mathrm{ej}}$, $m^{\mathrm{wind}}_{\mathrm{ej}}$, and $m^{\mathrm{dyn}}_{\mathrm{ej}}$ using these fits and marginalizing over \ac{EoS} can be seen in Fig. \ref{fig:dyn_wind} for a \ac{MDC} event with injected masses of $m_1 = 1.40 M_\odot$, ~$m_2 = 1.34 M_\odot$. Further discussion found in Sec. \ref{subsec:grid}.

\subsection{Inclination}
\label{subsec:inclination}

\begin{figure}
 \includegraphics[width=3.5in]{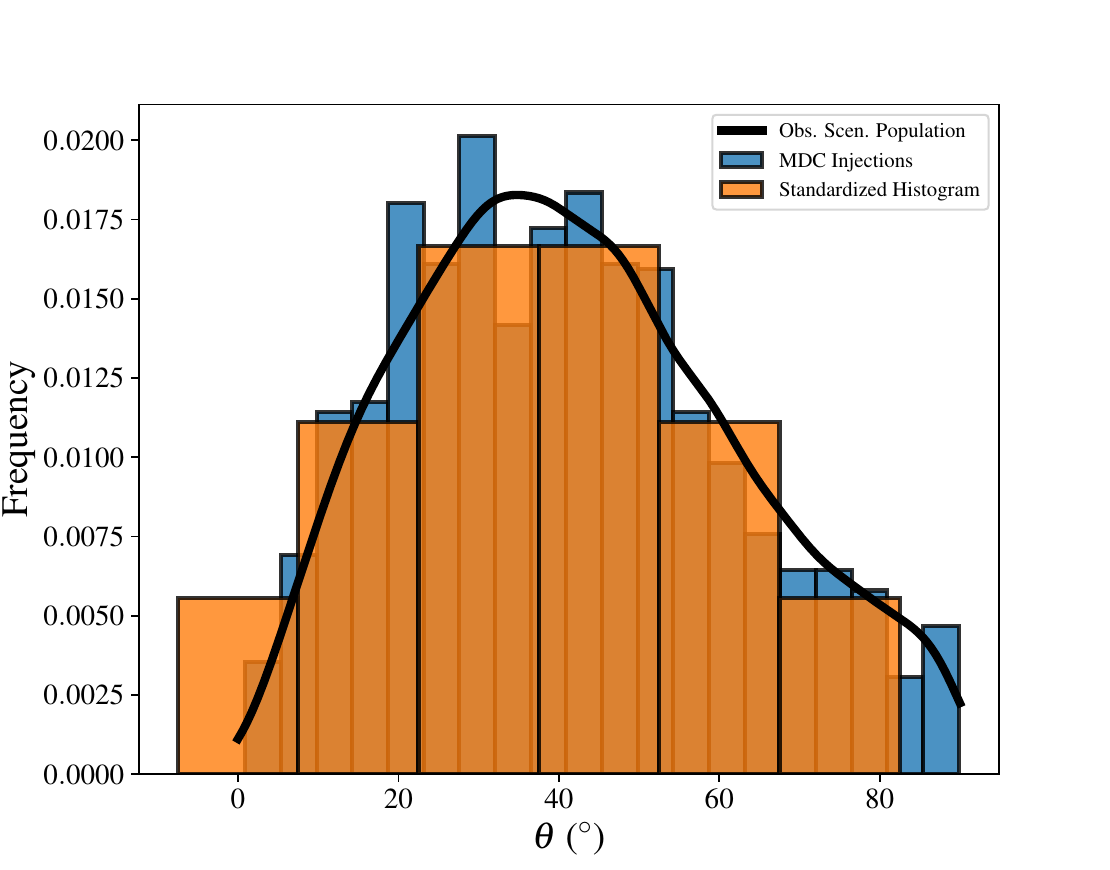}
  \caption{The two methods available for drawing $\theta$: the observing scenarios population distribution \citep{michael_w_coughlin_2022_7026209} (black) and a standardized histogram (orange), compared to the 
  histogram of $\theta$ from \ac{MDC} injection samples with predicted $ m_{\rm ej} \geq 10^{-3} M_\odot$ (blue). When using the observing scenarios distribution we draw a sample from the KDE for each initial component mass sample provided. When using the standardized histogram method, each initial component mass sample is run with $\theta = [0, 15, 30, 45, 60, 75]$ with counts proportional to the histogram, meaning each sample is run with the same $\theta$ values for consistency even with small sample sizes.}
 \label{fig:kde}
\end{figure}

The mass ejected from a neutron star is not thought to be perfectly uniform and spherical \citep{Heinzel:2020qlt, Sneppen:2023vkk}, which means the \ac{EM} emission from a kilonova will not be isotropic.
We expect to observe brighter emissions for face-on events, meaning $\theta \simeq 0^{\circ}$. In the case of kilonovae, this inclination angle, $\theta$, also known as the viewing angle, ranges from $0$ to $90^{\circ}$ as it is equivalent from $90$ to $180^{\circ}$ due to symmetry; in the GW case, face-on vs. face-off can be differentiated, and therefore measurements are reported from $0$ to $180^{\circ}$. Observationally, we would expect to see an $\theta$ distribution that is uniformly distributed in the projection along the line of sight, ($\cos{\theta}$).
However, because \ac{GW}'s have higher amplitudes for face-on events, we are in fact biased towards those face-on events, as those events will be louder and more likely to be detected. Taking both these factors into account, we will get a distribution that peaks at lower $\theta$ values and tails off toward higher $\theta$ values, as seen in Fig. \ref{fig:kde}. 

In the workflow, we have two means for drawing $\theta$. The first is shown by the black line in Fig. \ref{fig:kde}. This shows the $\theta$ distribution resulting from observing scenarios simulations \citep{michael_w_coughlin_2022_7026209, Kiendrebeogo:2023hzf}; these are based on GWTC-3 \citep{KAGRA:2021vkt} catalog fits which are simulated using \texttt{Bayestar} \citep{SiPr2016}, which simulates the matched filtering process and Gaussian noise in order to mimic realistic \ac{GW} detections. Fig. \ref{fig:kde} shows the KDE that estimates the $\theta$ distribution based on those simulations, and we can sample from this distribution. The second method uses standardized draws to mimic that same distribution, but instead of providing one random value, this method covers a range of $\theta$ values. For each sample, we cover the values 0, 15, 30, 45, 60, and 75 with probabilities proportional to what is seen in Fig. \ref{fig:kde}. This method removes the random nature of the draws which can be useful for small sample sizes. These two methods converge for large sample sizes and can be chosen based on the sample size and any time or resource constraints.

\subsection{Kilonova light curve model}
\label{subsec:lightcurves}

\begin{figure*}
 \includegraphics[width=\textwidth]{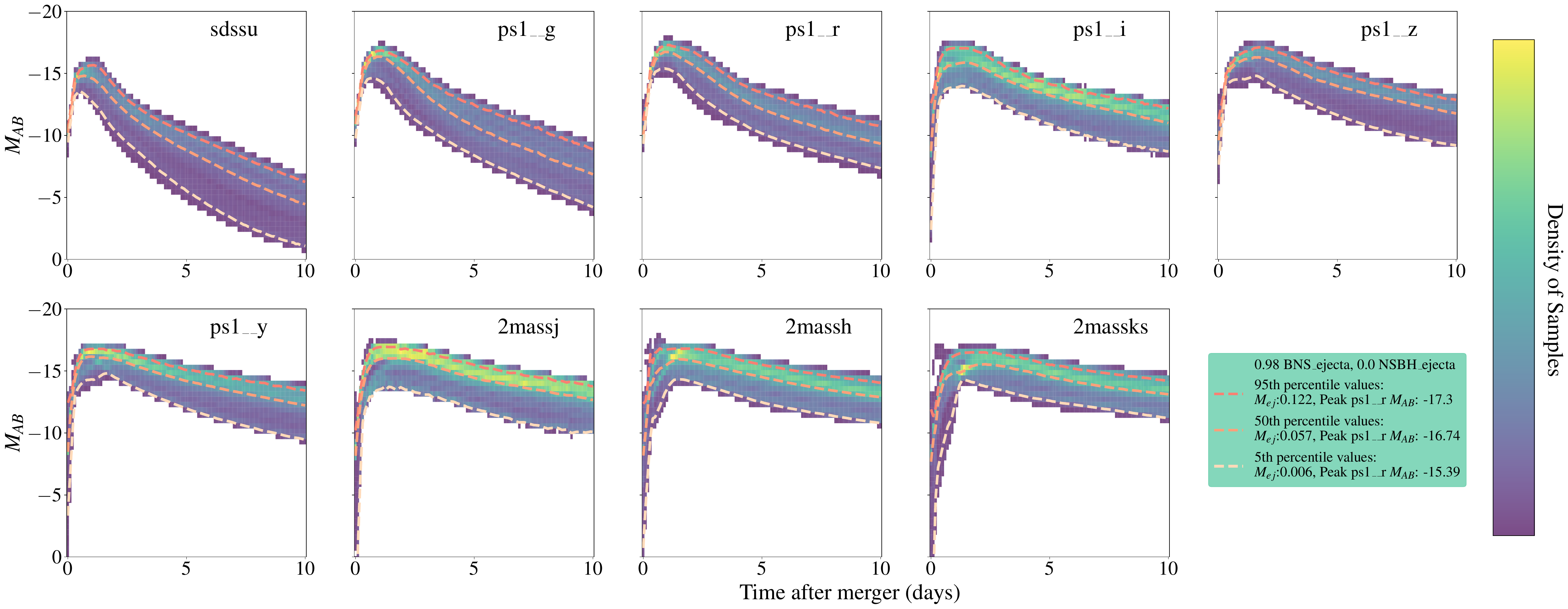}
  \caption{Heat map of light curves produced from \ac{EoS} marginalization on parameter estimation an \ac{MDC} event injected with a primary mass of $1.40  M_\odot$, and a secondary mass of $1.34  M_\odot$. The 5th, 50th, and 95th percentile predictions are shown by the colored dotted lines and their corresponding $m_{\mathrm{ej}}$ values can be found in the legend.} 
 \label{fig:heatmap1}
\end{figure*}

%\begin{figure*}
% \includegraphics[width=\textwidth]{plots/heatmap_lc_table_gp_1000_82x50_0.792.pdf}
%  \caption{Heat map of light curves produced from \ac{EoS} marginalization on parameter estimation an \ac{MDC} event injected with a primary mass of $1.87  M_\odot$, and a secondary mass of $1.43  M_\odot$. The 5th, 50th, and 95th percentile predictions are shown by the colored dotted lines, and their corresponding $m_{\mathrm{ej}}$ values can be found in the legend.}
% \label{fig:heatmap2}
%\end{figure*}

Our kilonova light curve model is based of the time-dependent, three-dimensional Monte Carlo radiative transfer code \texttt{POSSIS} \citep{Bul2019}, which can simulate kilonova light curves for a range of input parameters. A grid of these simulated kilonovae is then used to train a model in \ac{NMMA} in order to generalize predictions to arbitrary parameters. The model takes parameters of inclination angle ($\theta$), half-opening angle ($\phi$), dynamical ejecta $m^{\mathrm{dyn}}_{\mathrm{ej}}$, and wind ejecta $m^{\mathrm{wind}}_{\mathrm{ej}}$, and outputs light curves in nine \texttt{ugrizy} and \texttt{HJK} absolute magnitude ($M_{AB}$) bands. 
For this analysis, we have fixed this half-opening angle value to $30^{\circ}$, which was shown to be a reasonable fit to GW170817 \citep{Pang:2022rzc}. 

Fig.~\ref{fig:dyn_wind} shows a peak $r$ band $M_{AB}$ light curve predictions from a single simulated event with $m_1 = 1.40 M_\odot$ and $m_2 = 1.34 M_\odot$, and the associated $m_{\mathrm{ej}}$ distributions.
Fig.~\ref{fig:heatmap1} shows all light curves across nine bands for that same event, as well as the median and $90\%$ credible interval for $m_{\mathrm{ej}}$ and $M_{AB}$.
The right panel of Fig.~\ref{fig:scatter} shows peak $r$-band $M_{AB}$ across a range of component masses consistent with \ac{BNS} and \ac{NSBH} mergers. The points show the median value across \ac{EoS} realizations. In general, we find low mass \ac{BNS} events produce the brightest events, and $M_{AB}$ is strongly correlated with $m_{\mathrm{ej}}$. The light curve data products will be covered in detail and analyzed in Sec. \ref{sec:analysis}.

% covered above
%\subsection{light curves - surrogates}

%Light curves simualted with Possis, gird, neural net model in NMMANext we move onto the light curve models, which are highly dependent on $m_{\mathrm{ej}}$. We use the Bulla \citep{Bul2019} light curve model for light curve predictions. This model depends on the total $m_{\mathrm{ej}}$, the inclination angle, and the opening angle. We also have the ability to run predictions for different models, including Kasen et al \citep{KaMe2017}. All light curve predictions are done for the absolute magnitude in the source frame. If needed, masses are shifted from the detector frame using the redshift.

\section{Analysis of data products}
\label{sec:analysis}

\subsection{Proposed data products}

%We have particular interest in those that may produce kilonovae, and we look to not only estimate whether or not that may be possible due to a significant amount of mass ejected by the merger, but also how bright such kilonova light curves would be. We focus on predictions of the amount of mass ejected from the merger, and the peak magnitude of the possible kilonova. 

The first of our data products consists of three categories, which sum to a probability of 1: \ \texttt{BNS\_ejecta}, \texttt{NSBH\_ejecta}, and \texttt{no\_ejecta}. \texttt{BNS\_ejecta} refers to the probability that there will be significant $m_{\mathrm{ej}}$, defined as greater than $10^{-3} M_\odot$, produced by a \ac{BNS} merger. Similarly, \texttt{NSBH\_ejecta} refers to the probability that there will be significant $m_{\mathrm{ej}}$ produced by a \ac{NSBH} merger; finally, \texttt{no\_ejecta} refers to the probability that there will be $m_{\mathrm{ej}} < 10^{-3} M_\odot$. Each of these quantities assume the event is astrophysical in nature, and \texttt{no\_ejecta} is indiscriminate of the type of merger. Due to \ac{EoS} and parameter uncertainties, it is possible for predictions to contain non-zero values for both \texttt{BNS\_ejecta} and \texttt{NSBH\_ejecta}. Given \ac{EoS} realizations define the maximum mass of the neutron star, a sample near the border of a neutron star or black hole may have individual \ac{EoS} realizations classifying the object nature differently. The sum of \texttt{BNS\_ejecta} and \texttt{NSBH\_ejecta} will then provide the probability that a given event has significant ejecta, i.e. greater than $10^{-3} M_\odot$. We denote this quantity \texttt{HasEjecta}, and expect this to be useful metric for astronomers. These quantities specify likelihood of significant $m_{\mathrm{ej}}$, and are directly correlated to the likelihood a detectable kilonova exists, but we also take this one step further. 

In addition to the above data products, we provide $m_{\mathrm{ej}}$ and peak magnitude estimates in the form of three percentile values that constitute a median value and 90\% credible interval: the 5th, 50th, and 95th percentiles. We also create probability density maps of the light curves produced, showing the range of possible outcomes in a informative way, along with the percentile estimates (as seen in Fig.~\ref{fig:heatmap1}).

These complete $m_{\mathrm{ej}}$ and light curve predictions can be seen for an example \ac{MDC} event in Fig.~\ref{fig:heatmap1}. Shown are probability density maps for light curves found by starting with parameter estimation and marginalizing over \ac{EoS}. This event is entirely dominated by samples consistent with a \ac{BNS}: Fig.~\ref{fig:heatmap1} shows an event with a mass ratio near one ($m_1 = 1.40  M_\odot$, $m_2 = 1.34  M_\odot$). We also show the median and $90\%$ credible interval predictions for both $m_{\mathrm{ej}}$ and the peak $r$ band magnitude. These light curve predictions across \texttt{ugrizy} and \texttt{HJK} bands are intended for astronomers to help with \ac{EM} follow-up by providing an estimate of whether we can expect to see a kilonova, and how bright the kilonova may be.

\subsection{Predictions compared to GW170817 + AT2017gfo}

%We see that in both cases, the injection falls well with the 90\% credible interval. 

\begin{figure}
 \includegraphics[width=3.5in]{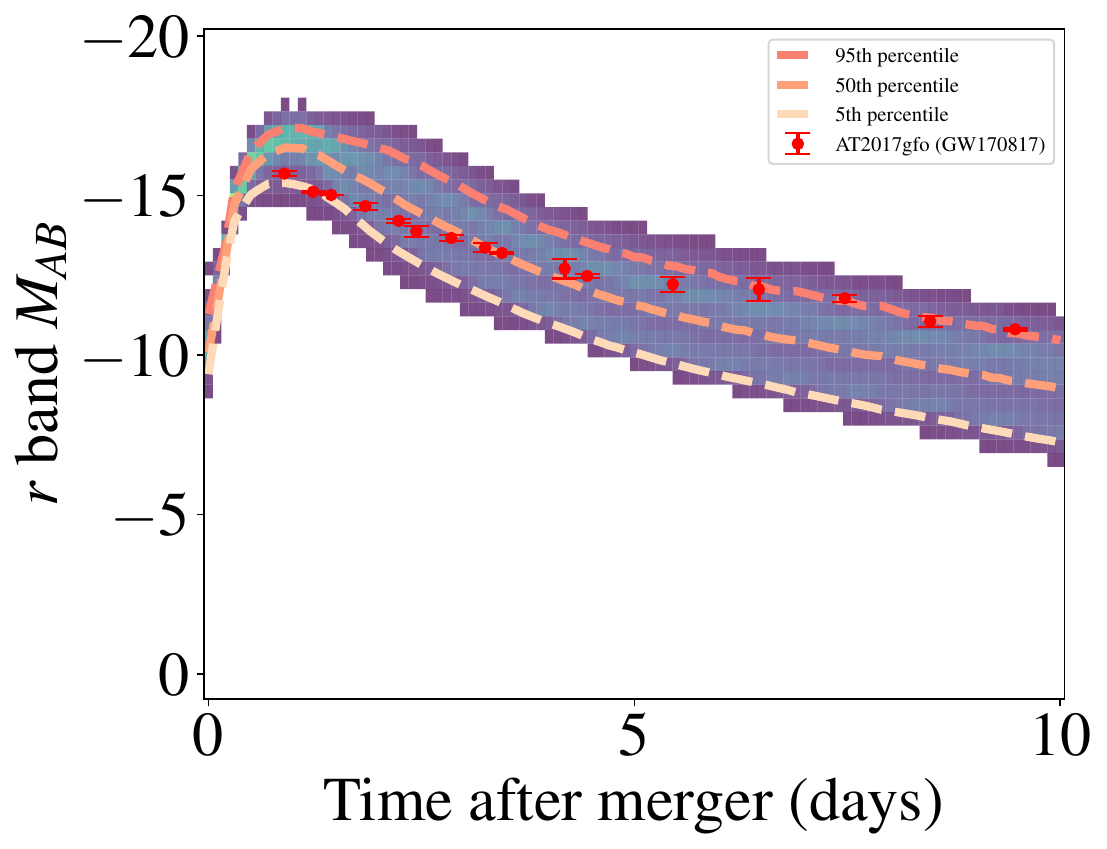}
  \caption{Comparison of light curve predictions using GW170817 posterior samples to At2017gfo light curves. The observed $r$ band light curves fall within the $90\% $ credible interval of the predicted light curves.}
 \label{fig:GW170817}
\end{figure}

As a sanity check for our light curve predictions, we compare to AT2017gfo \citep{CoFo2017,SmCh2017,AbEA2017f} the kilonova resulting from the \ac{BNS} GW170817 \citep{AbEA2017b}. We expect the observed light curves to fall within the $90\%$ credible interval of our prediction. To make predictions, we start with parameter estimation from GW170817, and run the predictions workflow including \ac{EoS} marginalization, producing a range of light curves, as seen in Fig. \ref{fig:GW170817}. We focus on the $r$ band $M_{AB}$ for the sake of this plot, and we find that the observed data does in fact fall within our $90\%$ credible interval of predictions as expected.

\subsection{Predictions across a grid of component masses}
\label{subsec:grid}

\begin{figure*}
 \includegraphics[width=3.4in]{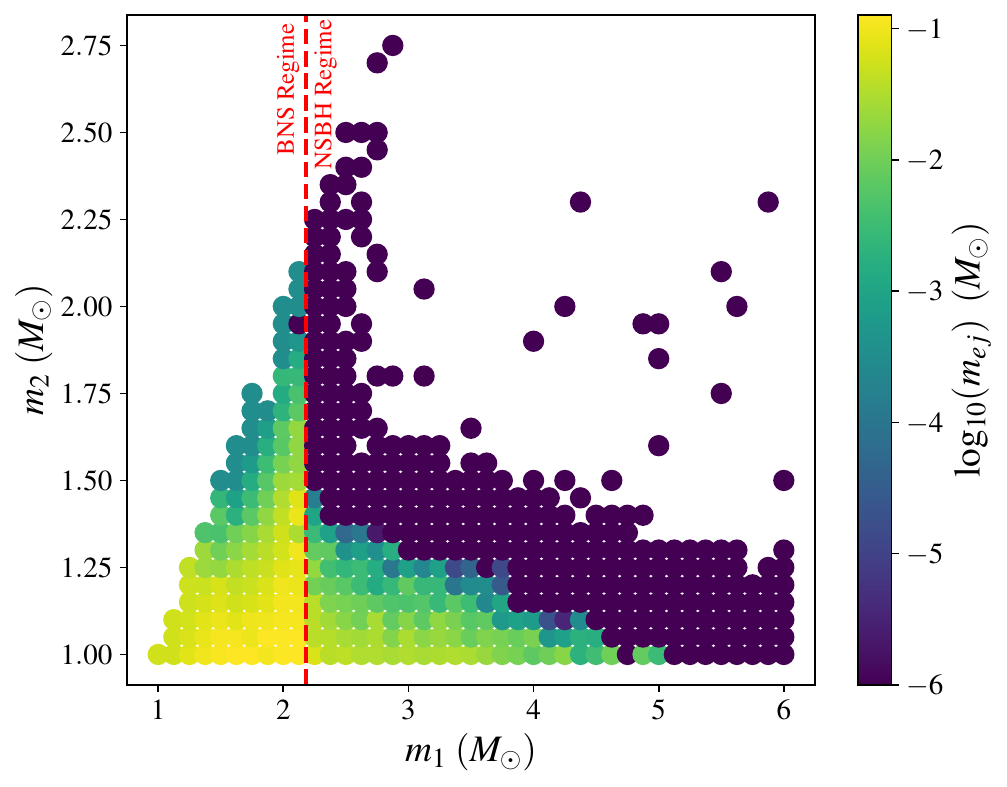}
 \includegraphics[width=3.5in]{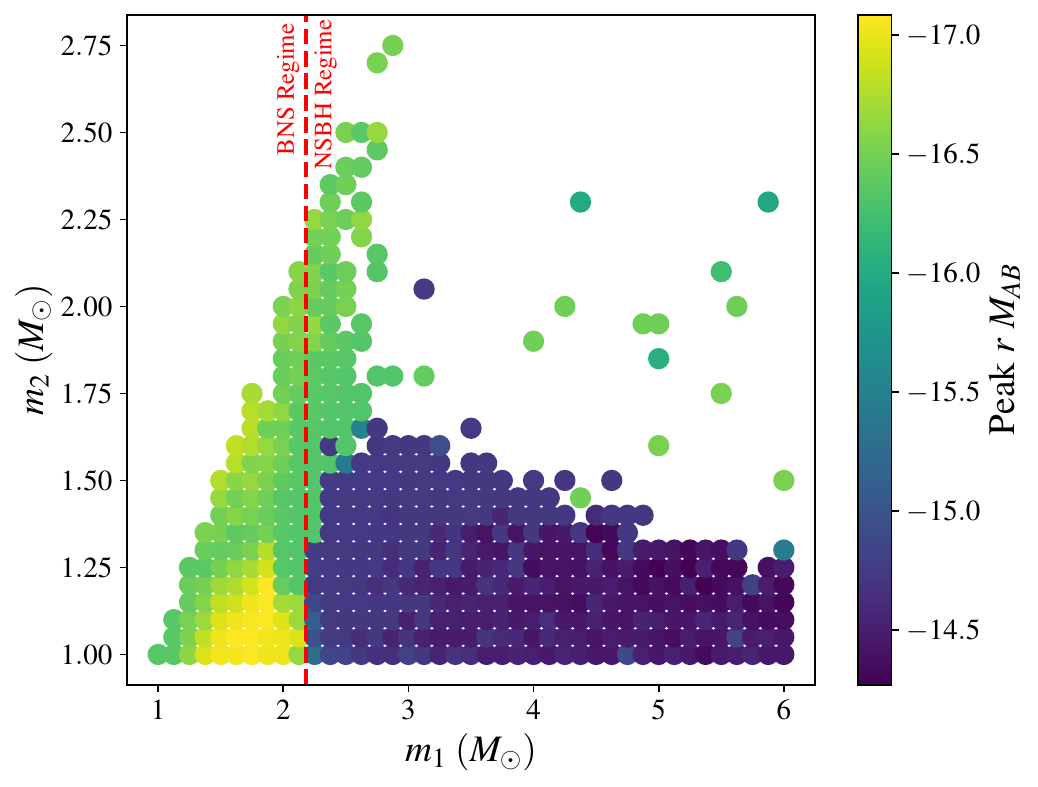}
  \caption{\textit{Left:} A scatter colored by the median $m_{\mathrm{ej}}$ produced for a given component mass pair, marginalized over a number of \ac{EoS} realizations. This plot only shows points that have some samples greater than $10^{-3} M_{\odot}$ and the low end of the color bar is capped at $10^{-6} M_\odot$. The \ac{BNS} and \ac{NSBH} regimes are indicated by a dotted line, loosely defined by a maximum \ac{NS} mass of $\simeq$ 2.15 $M_{\odot}$. We find the low mass \ac{BNS} regime produces the largest $m_{\mathrm{ej}}$, and see a significant drop off for the \ac{NSBH} regime. \textit{Right:} A the scatter colored by the median $r$ band $M_{AB}$ for a given component mass pair. The distribution is similar to the left panel due to the correlation of $m_{\mathrm{ej}}$ with the brightness of the kilonova, and with fainter light curves in the \ac{NSBH} regime due to the different light curve model. For the consistency of the plot, all spins are set to zero and $\theta$ is drawn from the histogram method in Sec. \ref{subsec:inclination}.
  }
 \label{fig:scatter}
\end{figure*}

In order to provide an idea of what regimes in the parameter space are most likely to produce an observable kilonova, we make predictions across a wide grid of component masses spanning \ac{BNS} and \ac{NSBH} events. For the consistency of this plot we assume zero spin for all objects. The left panel of Fig.~\ref{fig:scatter} shows the median $m_{\mathrm{ej}}$ predictions across \ac{EoS} realization for a grid of \ac{BNS} and \ac{NSBH} mergers. The dotted line between the \ac{BNS} regime and \ac{NSBH} regime denotes where the two source classifications dominate. Each sample is classified as \ac{BNS} or \ac{NSBH} based on the maximum \ac{NS} mass of the \ac{EoS} realization, and use the appropriate fits and models. We find the the low mass \ac{BNS} events have the largest $m_{\mathrm{ej}}$, and that \ac{NSBH} events tend to have less $m_{\mathrm{ej}}$. The same trends follow for peak $r$ band $M_{AB}$, with a sharp drop off as we move from the \ac{BNS} to the \ac{NSBH} light curve model, again pointing to the fact \ac{BNS} mergers are most likely to produce an observable kilonovae.

%\rednote{maybe consider showing the same plots for spin=1 in an Appendix}.

\subsection{Predictions from Mock Data Challenge events}

In order to validate our proposed data products, we use \ac{MDC} \citep{Meacher:2015iua, Chaudhary:2023vec} events. The \ac{MDC} is a real-time simulation campaign where \ac{CBC} waveforms are injected into \ac{O3} strain data. \ac{CBC} searches are carried out and event candidates are uploaded internally to \ac{GraceDB}\footnote{\url{https://gracedb.ligo.org/}}, including downstream data products such as parameter estimation and sky localizations. This provides us with a set of parameter estimation posterior samples for \ac{CBC} events, and also allows us to refer back to the original simulated quantities for comparison. 

We produced our data products for a set of \ac{MDC} events, including both \ac{BNS} and \ac{NSBH}, for which we predict the likelihood of generating $m_{\mathrm{ej}}$, as well as the brightness of the resulting light curves.
As described in the workflow above, we use the masses and spins generated by the parameter estimation, while marginalizing over \ac{EoS}, to make the predictions. 
In order to compare our predictions to the injections and to sanity check our results, we run a set of \ac{MDC} events with our \ac{EoS} marginalized predictions, and compare those to predictions made using the injections and the SLy \ac{EoS}.

Fig. \ref{fig:dyn_wind} shows a detailed view of how $m^{\mathrm{dyn}}_{\mathrm{ej}}$, $m^{\mathrm{wind}}_{\mathrm{ej}}$, and $m_{\mathrm{ej}}$ vary across parameter estimation posterior samples from a single simulated event. We find all four quantities are highly correlated, with the highest $m_{\mathrm{ej}}$ and brightest $r$ band $M_{AB}$ values corresponding to small secondary \ac{NS} mass in this case.

%Fig.~\ref{fig:MDCpercentiles} plots the percentile that the SLy injection run falls within the \ac{EoS} marginalized samples for $m_{\mathrm{ej}}$. We find that our \ac{EoS} marginalized predictions tend to slightly overestimate $m_{\mathrm{ej}}$ for lower masses, and slightly underestimate for higher masses compared to the injection run with SLy. This is because the SLy \ac{EoS} is just an example (but reasonably plausible) \ac{EoS}; however, we see that the vast majority of events will fall within the $90\%$ credible interval we will report, showing that the recovery is reasonable.

%\begin{figure}
% \includegraphics[width=3.5in]{plots/combined_plot.pdf}
%  \caption{Top: Scatter of \ac{BNS} events recovered in the MDC. We predict the $m_{\mathrm{ej}}$ using the \ac{MDC} injection using the SLy \ac{EoS} and using the full set of posterior samples and \ac{EoS} marginalization for the same events. We then find the percentile of the injection value relative to the full \ac{EoS} predictions, and color the points by those percentiles. We find our predictions tend to slightly overestimate $m_{\mathrm{ej}}$ for lower masses, and slightly underestimate for higher masses. Bottom: Histogram of those same percentile values comparing the injected predictions to our \ac{EoS} marginalized ones. We find a mostly uniform spread of percentiles.}
% \label{fig:MDCpercentiles}
%\end{figure}

\begin{figure*}
     \includegraphics[width=\textwidth ]{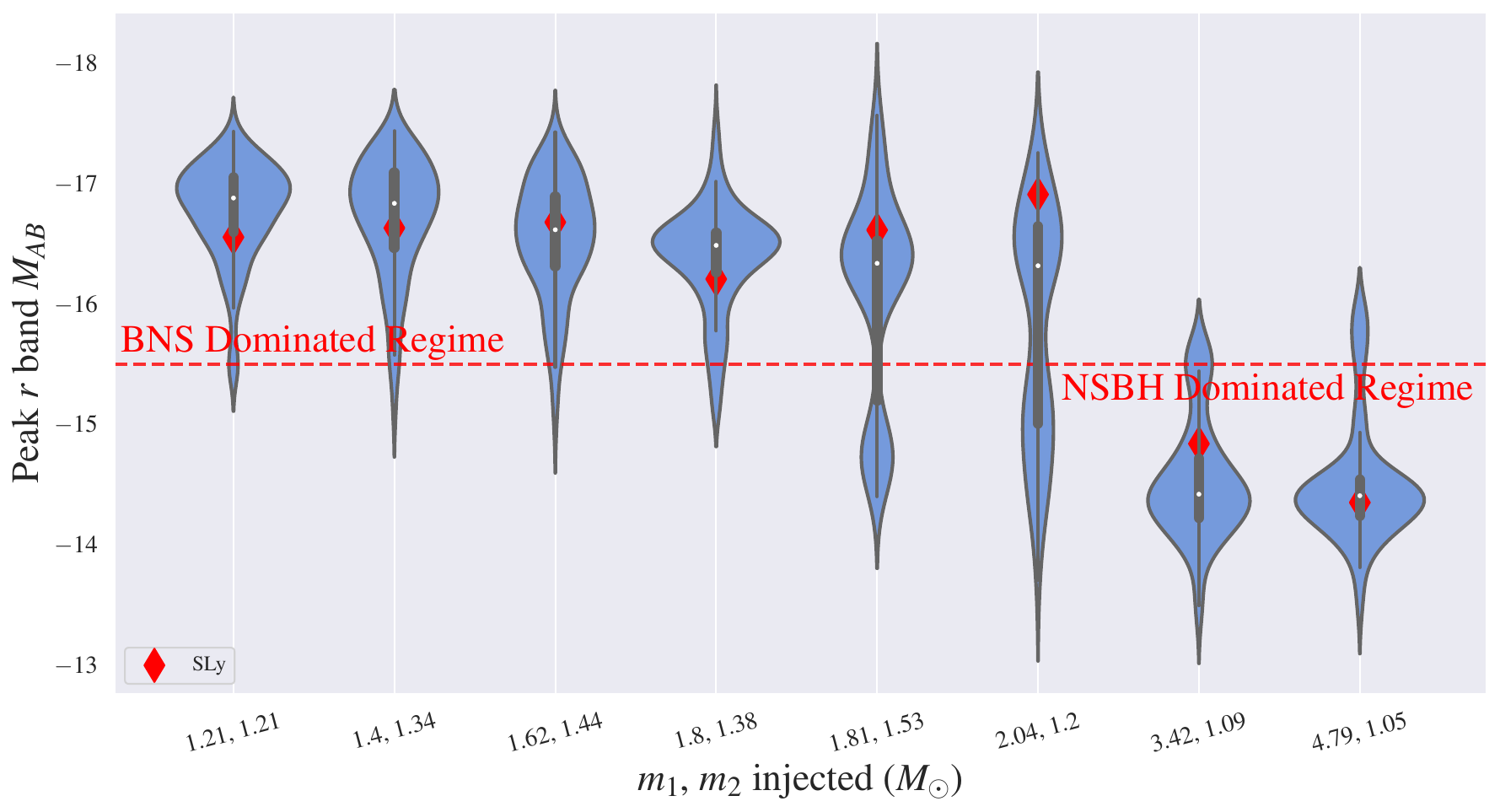}
  \caption{Violin plot showing peak $r$ band $M_{AB}$ predictions for 8 \ac{MDC} events of increasing total mass. The range of predictions found from parameter estimation samples are compared to the injected parameters run with the SLy \ac{EoS}. The curves show a KDE of the samples while inside the colored region there is a box and whisker plot. A dotted line is used to indicate the region dominated by \ac{BNS} and \ac{NSBH} samples, respectively.}
 \label{fig:violin}
\end{figure*}

Further, Fig.~\ref{fig:violin} shows a violin plot of how the \ac{EoS} marginalized predictions for parameter of estimation \ac{MDC} events of increasing total mass compare to one another and their injected source properties run with the SLy \ac{EoS}. These events are meant to roughly cover a range component mass pairs that may be capable of producing a kilonova. As with all predictions in this paper, the distributions shown here are only for samples with $m_{\mathrm{ej}} \geq 10^{-3} M_\odot$. The curves surrounding the colored regions are KDE approximations of the samples, with a box and whisker plot enclosed and the median represented by the white dot. We find that the \ac{EoS} marginalized distributions and the 90\% credible intervals are generally consistent with the injections run with SLy. 

Moving from left to right with increasing total mass, we find the median peak $r$ band $M_{AB}$ values follow a general decreasing trend, consistent with what is seen in Fig. \ref{fig:scatter}. The red dotted line shows an approximate break between the \ac{BNS} and \ac{NSBH} regimes, where separate ejecta fits and light curve models dominate. This is the reason we see the bimodality of the violin plots, with the \ac{BNS} samples mostly concentrated on top, and the \ac{NSBH} samples on bottom. This is also obvious for the SLy injected values, as there is a significant drop off between the first six events (classified as \ac{BNS} by SLy), and the last two (classified by \ac{NSBH} by SLy), as we cross the SLy maximum \ac{NS} mass of $\simeq 2.1 M_\odot$. A single event can have samples in both the \ac{BNS} and \ac{NSBH} regime as parameter estimation covers a range of mass pairs, mostly consistent with a well constrained chirp mass value but with varying mass ratio, and that our range of \ac{EoS}s that we marginalize over each define their own maximum \ac{NS} mass. In this way, we show our range of predictions including these uncertainties, and across source classifications. This plot shows \ac{BNS} are most likely to produce bright, observable kilonovae based on our current understanding of the underlying factors.

Finally, we note that these representations may be an additional way of viewing the overall distribution, and could be produced for \ac{GW} candidates.

\section{Conclusion}
\label{sec:conclusion}

In this paper, we propose new kilonova light curve and ejecta mass data products to help inform electromagnetic follow-up of \ac{GW} candidates from \ac{CBC} searches. 
To make predictions, we marginalize over \ac{EoS}, and use ejecta fits and a light curve model to produce out estimates. Our data products include the probabilities \texttt{HasEjecta}, which describes the probability of an candidate event having a $m_{\mathrm{ej}} \geq 10^{-3} M_\odot$, and \texttt{BNS\_ejecta} and \texttt{NSBH\_ejecta}, which describe the probability of \ac{BNS} and \ac{NSBH} mergers having $m_{\mathrm{ej}} \geq 10^{-3} M_\odot$. We also produce 90\% percent credible interval estimates for $m_{\mathrm{ej}}$, and \texttt{ugrizy} and \texttt{HJK} $M_{AB}$ bands.

We find that our predictions are not only consistent with GW170817 and AT2017gfo in Fig. \ref{fig:GW170817}, but also with injections run with the SLy \ac{EoS} in Fig. \ref{fig:violin}. Additionally, Fig. \ref{fig:violin} demonstrates \ac{BNS} are most likely to produce a bright, observable kilonova, and points to an inverse correlation between median peak $r$ band $M_{AB}$ total mass.

As we intend for these quantities to inform follow-up decisions and help enable astronomers to detect future kilonovae, we will advocate to make the following ejecta and light curve data products public during the second half of \ac{O4}. Our data products will not only help determine the likelihood that a \ac{GW} event could produce a kilonova, but also estimate the amount of $m_{\mathrm{ej}}$ and the light curves produced, the direct observables astronomers need when making follow-up decisions.

\emph{Acknowledgements.} 
We thank Naresh Adhikari for review of this paper. This material is based upon work supported by NSF's LIGO Laboratory which is a major facility fully funded by the National Science Foundation. The authors are grateful for computational resources provided by LIGO
Laboratory and are supported by NSF Grants No. PHY-0757058 and No.
PHY0823459. This material is based upon work supported by NSF’s LIGO
Laboratory, which is a major facility fully funded by the National
Science Foundation. This work used Expanse at the San Diego Supercomputer Cluster through allocation AST200029 -- ``Towards a complete catalog of variable sources to support efficient searches for compact binary mergers and
their products'' from the Advanced Cyberinfrastructure Coordination
Ecosystem: Services \& Support (ACCESS) program, which is supported by
National Science Foundation grants \#2138259, \#2138286, \#2138307,
\#2137603, and \#2138296. AT and MWC acknowledge support from the National Science Foundation with grant numbers PHY-2308862 and PHY-2117997. HG, JM, and ST acknowledge support from the University of Minnesota's UROP program. SSC and MC acknowledge support from the National Science Foundation with grant number PHY-2011334 and PHY-2308693. MC would also acknowledge support from NSF PHY-2219212. DC would like to acknowledge support from the NSF grants OAC-2117997 and PHY-1764464. GM acknowledges support from the National Science Foundation with grant number PHYS-2012017. T.D. acknowledges support by the European Union(ERC, SMArt, 101076369). Views and opinions expressed are those of the authors only and do not necessarily reflect those of the European Union or the European Research Council. Neither the European Union nor the granting authority can be held responsible for them. PL and RE are supported by the Natural Sciences \& Engineering Research Council of Canada (NSERC).

\bibliographystyle{mnras}
\bibliography{refs.bib} % if your bibtex file is called example.bib

%%%%%%%%%%%%%%%%%%%%%%%%%%%%%%%%%%%%%%%%%%%%%%%%%%

% Don't change these lines
\bsp	% typesetting comment
\label{lastpage}
\end{document}